# High Success Probability, Fidelity, and Purity Nonlinear Optical Two-Qubit Gates on Chip

Minghao Shang[1,†], Hua-Ying Liu[1,†,‡], Ying Wei[1], Xiaoyi Liu[1], Qianhao Ning[2], Lijian Zhang[1], Shi-Ning Zhu[1], and Zhenda Xie[1,‡]

[1]National Laboratory of Solid State Microstructures, School of Electronic Science and Engineering, School of Physics, College of Engineering and Applied Sciences, and Collaborative Innovation Center of Advanced Microstructures, Nanjing University, Nanjing 210093, China

[2]Harbin Institute of Technology, Harbin 150001, China

[†]These authors contributed equally to this work.

[‡]Contact authors: liuhuaying@nju.edu.cn, xiezhenda@nju.edu.cn

**Abstract:**

Optical two-qubit gate with high success probability, fault-tolerant fidelity, and high-purity outputs is a fundamental yet unsolved challenge, essential for large-scale optical quantum computing toward quantum advantage. Here, we propose a feasible scheme for such gate using thin-film lithium niobate platform, enabling $\chi^{(2)}$ nonlinear photon-photon interaction with 100% efficiency. By decoupling photon interaction and qubit flip operations, fidelity ceiling is removed, and output state purity is recovered by spectral-phase pre-compensation based on a full-spectral photon interaction model, yielding a CNOT gate with 84% success probability, 93% purity, and unity fidelity.

**Main text**

Photons are attractive carriers for quantum information [1-3], as their weak interaction with the environment gives rise to long coherence times and enables quantum operation even at room temperature. Yet its potential in large-scale quantum computing towards quantum computational advantage has been hindered, because of the challenge in development of scalable two-qubit optical logic gates. High success probability has to be achieved, as the overall computing success rate will decrease exponentially as the system scales up. Fidelity above the 99% level is also necessary, as its the threshold for fault-tolerant quantum computing [4-8]. High output-state purity is also necessary, because the indistinguishability of the photons determines their visibility for following interference operations [9,10]. So far, though two-qubit optical logic gates have been demonstrated in a variety of linear-optics architectures [9,11,12], their intrinsic probabilistic nature remains a central obstacle to large-scale operation.

On the other hand, nonlinear optics offers a fundamentally different route, which can enable logic gates with unity success probability in principle, if nonlinear photon-photon interaction can be achieved with 100% efficiency [13-15]. Recently, with the advance in integrated optics platforms with strong nonlinearity [16-21], the possibility for such interaction has emerged. Typically, theoretical work has proven that in the thin-film lithium niobate (TFLN) microresonator [16], single-photon parametric down conversion and two-photon up conversion can be achieved with unity efficiency. However, existing nonlinear optical gate framework follows the controlled-phase principle [13-15,22], which intrinsically results in reduced fidelity and purity, creating the bottleneck of scalability. Specifically, as the logical operation is encoded directly in a frequency-dependent nonlinear phase shift, the gates are intrinsically spectral dependent and therefore suffer from a fidelity ceiling for photons with bandwidth determined by the same process [22]. Besides, the nonlinear conversion process will inevitably introduce spectral correlations between the output photons, which degrades their purity for the following operations [10,23]. Hence, a new scheme is needed to avoid and suppress the spectral-phase-correlation-induced fidelity limitation and purity degradation, respectively, while maintaining a high success probability at the same time.

Here we propose a feasible scheme for nonlinear two-qubit gates, typically the controlled-NOT (CNOT) on chip, which can achieve high success probability, fault-tolerant fidelity, and high-purity outputs at the same time. Unlike the controlled-phase principle, the nonlinear interaction is used only for conditional photon conversion, while the logical qubit flip is implemented in a separate linear operation, as shown in

Fig. 1. In this way, the logical operation is decoupled from the frequency-dependent nonlinear phase, so the intrinsic fidelity ceiling of conventional controlled-phase gates can be removed. To recover the output state purity, we incorporate spectral-phase pre-compensation during linear operation to suppress the spectral correlation generated in the nonlinear stage. Implementing the scheme in the integrated TFLN platform, we provide a circuit architecture combining periodically poled microring resonators, wavelength-selective router, and electro-optic modulaters [17,18,24,25], for nonlinear photon-photon interaction, qubit flip, and electro-optic spectral-phase pre-compensation, respectively. Specifically, we develop a full spectral scattering treatment to evaluate the state involvement in the microresonator based on the Lippmann-Schwinger formalism [26-30]. Based on such model, we achieve an optimized gate design with an overall success probability of 84.1%, allowing unity fidelity in principle, and yields an output-photon purity of 93.1%, with required microring quality factors ($Q$) on a feasible order of only $10^7$. Such a scalable two-qubit gate on chip is a promising building block for large-scale integration, which fills in the missing puzzle for complex optical quantum computing towards many practical quantum tasks.

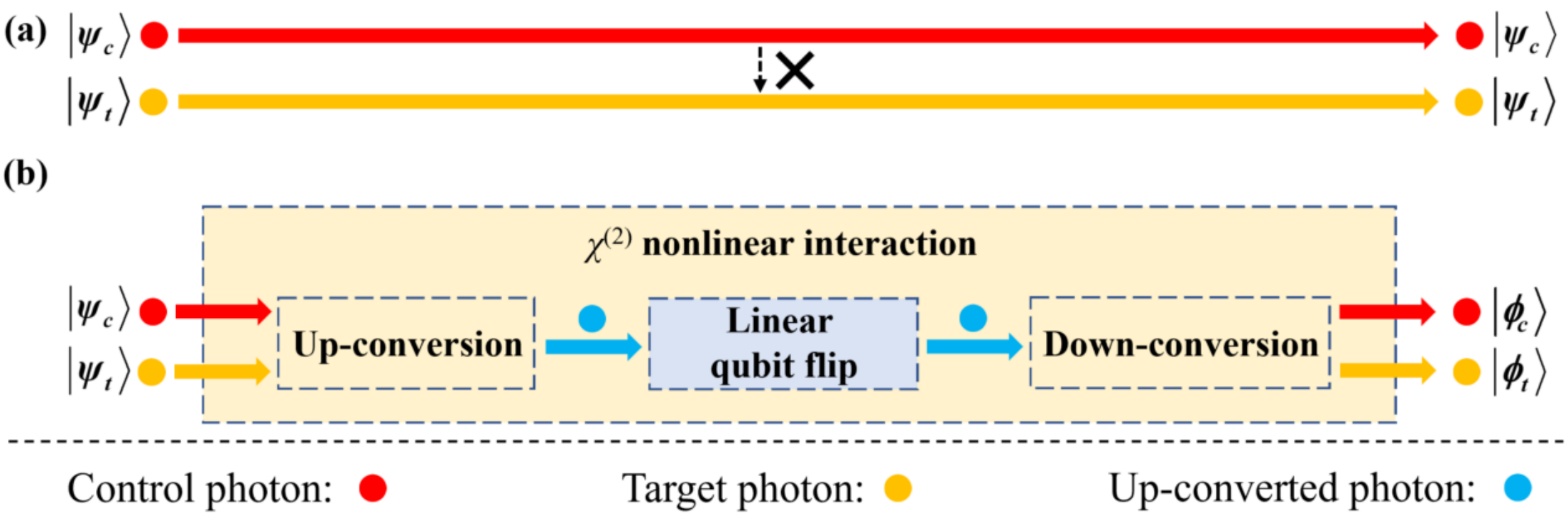

Fig. 1. Schematic nonlinear optical two-qubit gate operation based on $\chi^{(2)}$ nonlinearity and linear operations. (a) When optical nonlinearity is not introduced, there is no interaction between two photons. (b) By introducing strong nonlinear photon-photon interactions (including two-photon parametric up conversion and single-photon parametric down conversion) with single-photon linear operations, qubit flips with one photon controls another can be achieved. This decouples the logical operation from the frequency-dependent nonlinear phase and enables unity fidelity in principle.

Our gate architecture is illustrated in Fig. 2(a). The gate operation is realized in path encoding. When the control qubit is in the path state $|\tilde{1}\rangle_c$, it conditionally enables nonlinear interaction with the target photon. Through strong $\chi^{(2)}$ photon-photon interaction, the control-target photon pair undergoes two-photon parametric up-conversion (TPPUC) and is converted into an auxiliary up-converted photon. The qubit flip is then implemented in the linear stage through path routing of the up-converted photon. Finally, the single-photon parametric down-conversion (SPPDC) process restores the up-converted single photon to a two-photon state, thereby realizing a CNOT gate operation [9],

$$\hat{\mathbf{U}}_{\text{CNOT}} = \begin{pmatrix} 1 & 0 & 0 & 0 \\ 0 & 1 & 0 & 0 \\ 0 & 0 & 0 & 1 \\ 0 & 0 & 1 & 0 \end{pmatrix}. \tag{1}$$

A central feature of this architecture is that the logical qubit flip is not performed by the nonlinear interaction itself, but by a separate linear stage. This operating principle differs fundamentally from the conventional nonlinear controlled-phase schemes, in which the gate operation is encoded in a nonlinear phase shift $\varphi$ [9,13],

$$\hat{\mathbf{U}}_{\text{CPhase}} = \text{diag}\left(1,1,1,e^{i\varphi}\right), \tag{2}$$

where the ideal $\varphi$ equals $\pi$, taking the controlled-Z (CZ) gate as an example. For realistic photon wave packets with certain bandwidth, however, the phase shift becomes frequency dependent, i.e., $\varphi(\omega_c, \omega_t)$, where $\omega_c$ and $\omega_t$ are the angular frequencies of control and target photons, respectively. As a result, the operation is no longer described by a single unitary acting only on the computational subspace. Instead, the frequency-dependent phase couples the computational and spectral degrees of freedom, so that the phase shift is no longer the constant value $\pi$, but a frequency-dependent function $\varphi(\omega_c, \omega_t)$. Tracing over the spectral degrees of freedom therefore yields an effective mixed operation in the computational space, giving rise to an intrinsic fidelity ceiling. Importantly, this limitation does not originate from insufficient nonlinear interaction strength, but from encoding the logical operation in a frequency-dependent nonlinear phase. While in our architecture, the nonlinear process is used only for conditional photon conversion. For this reason, the logical operation is decoupled from the frequency-dependent nonlinear phase during photon-photon interaction. In practice, the fidelity is then limited only by imperfections of linear devices, such as wavelength-division multiplexers (WDM) and propagation loss. Given the current maturity of integrated linear photonic devices [31,32], fidelities exceeding the fault-tolerant thresholds [4-8] are in principle accessible. In addition to removing such fidelity ceiling, the architecture should also address spectral correlations generated in the SPPDC process to preserve scalability. To this end, the linear stage is extended to a pre-compensated WDM (PC-WDM) module that combines the qubit flip with spectral-phase modulation of the up-converted photon. This allows suppression of the phase and amplitude correlations of the down-converted output state.

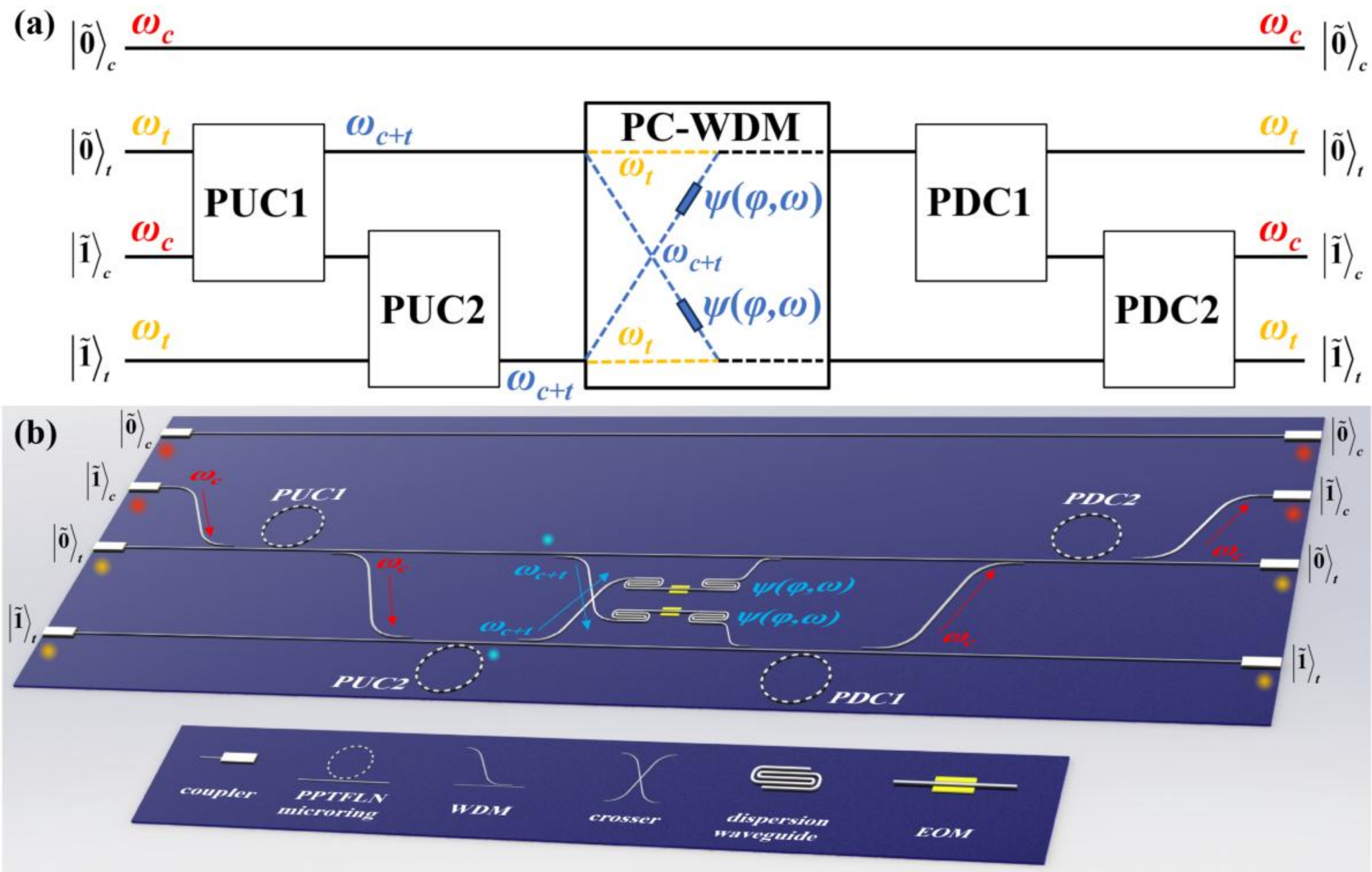

Fig. 2. (a) Schematic diagram of the efficient and unity-fidelity CNOT gate scheme based on $\chi^{(2)}$ integrated optics platform, using the non-degenerate control photon $\omega_c$ and the target photon $\omega_t$ to control the path encoding. The PC-WDM routes photons of the specified frequency to the corresponding paths, while pre-compensating the phase and modulus correlation induced by SPPDC process. (b) An on-chip scheme of the CNOT gate. PPTFLN microring resonators are used for TPPUC and SPPDC. The large-dispersion waveguides and EOMs are utilized for pre-compensation.

To implement the gate principle in a realistic integrated setting, we employ TFLN photonics [17,18,24,25]. This platform simultaneously provides strong $\chi^{(2)}$ nonlinearity for photon-photon interaction, mature linear routing components for path control, and outstanding electro-optic (EO) functionality for spectral-phase control, making it particularly suitable for the present architecture. The on-chip implementation is shown in Fig. 2(b). Cascaded TPPUC and SPPDC are realized in periodically poled TFLN (PPTFLN) microring resonators using quasi-phase matching (QPM) method [33,34]. The linear qubit flip is implemented by frequency-controlled routing with WDMs [32], while spectral-phase pre-compensation is implemented using dispersion waveguides and EO modulators (EOMs) [35]. In this way, the same integrated platform supports both the nonlinear conversion required for conditional qubit control and the linear operations required for removing the phase-induced fidelity limitation and mitigating output spectral correlations.

It should be emphasized that, to realize the cascaded TPPUC-SPPDC processes with high efficiency and spectral confinement, we employ microring resonators rather than straight waveguides. This choice is motivated by three considerations. First, cavity enhancement in the triply resonant configuration substantially strengthens the effective $\chi^{(2)}$ interaction, which is essential for achieving high nonlinear conversion efficiency at single-photon level [16]. Second, the compact footprint reduces sensitivity to fabrication nonuniformity and defects in the thin-film platform. Third, and most importantly for the present work, the narrow cavity resonances constrain the photon spectra and suppress spectral broadening during cascaded nonlinear processes, thereby preserving the narrow output spectra required for large-scale cascading.

Because the gate operates through spectrally relevant nonlinear conversion, the success probability and the output-state purity are both determined by the detailed spectral structure of the cascaded TPPUC and SPPDC processes. A quantitative treatment of the full spectral dynamics is therefore essential. For this reason, single-frequency approximation [14-16] is insufficient here, and we construct a full spectral scattering model for the triply resonant microring-bus system shown in Fig. 3.

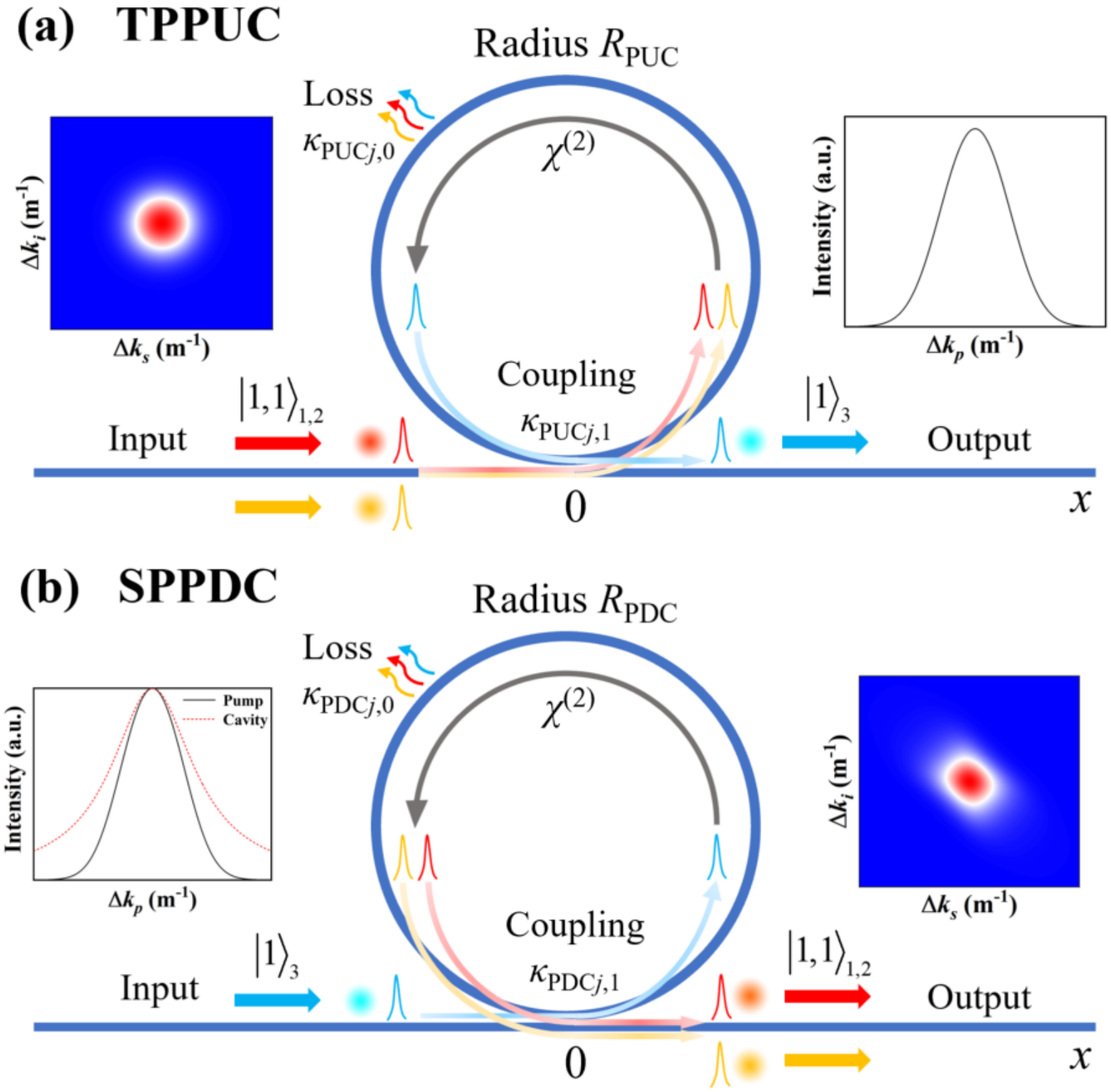

Fig. 3. The model of photon-photon interaction in TFLN microring resonators with bus waveguides for coupling, considering the spectra of input photons. Considering the coordinate representation, the coupling coordinate of the ring and the bus waveguide is set to be $x = 0$, while $x < 0$ and $x > 0$ represents coordinates before coupling and after coupling, respectively. (a) The TPPUC model, where the spectra of input photon pair are uncorrelated. (b) The SPPDC model. Typically, the output two-photon joint spectral intensity (JSI) shows a correlation.

In this model, the total Hamiltonian is written as

$$\hat{H}_{\text{system}} = \hat{H}_{\text{microring}} + \hat{H}_{\text{bus}} + \hat{H}_{\text{coupling}}, \tag{3}$$

where the three terms describe the discrete cavity modes [16], the continuous bus-waveguide modes [26], and their mutual coupling [26], respectively. In the common time-evolution approaches, the continuous-spectral Hilbert space leads to an infinite set of coupled equations, making the problem analytically intractable [36]. Instead, we employ the L-S scattering formalism [26-30], which allows direct calculation of the analytical output states without explicitly solving the full dynamical evolution. This approach is particularly suitable here because the input and output states are spatially separated from the interaction region. Moreover, the quantities of interest—conversion efficiency, joint spectral amplitude, and output-state purity—are naturally expressed in the scattering picture. The continuous-spectral Hamiltonian of the bus, expressed in coordinate space, is then given by [26]

$$\hat{H}_{\text{bus}} = \hbar \int \mathrm{d}x \sum_{j=1}^{3} v_{gj} \hat{b}_j^+(x) \frac{1}{i} \frac{\mathrm{d}}{\mathrm{d}x} \hat{b}_j(x), \tag{4}$$

where $v_{gj}$ ($j$ = 1, 2, 3) is the group velocity at the center angular frequencies of $\omega_1$, $\omega_2$, $\omega_3$, respectively, with $\omega_3 = \omega_1 + \omega_2$, and $\hat{b}_j(x)$ stands for the annihilation operator of photons in the bus waveguide. And the Hamiltonian of the coupling between the ring and the bus [26] is

$$\hat{H}_{\text{coupling}} = \sum_{j=1}^{3} \hbar V_j \int \mathrm{d}x \delta(x) \left( \hat{b}_j^+(x) \hat{\varsigma}_j + \hat{b}_j(x) \hat{\varsigma}_j^+ \right), \tag{5}$$

where $V_j$ is the coupling strength, and $\hat{\varsigma}_j$ stands for the photon annihilation operator in the microring resonator. On the other hand, the Hamiltonian of the microring is in the discrete-spectral Hilbert space, which can be expressed in the interaction picture:

$$\hat{H}_{\text{microring}} = \hbar \xi \left( \hat{\varsigma}_3^+ \hat{\varsigma}_1 \hat{\varsigma}_2 + \hat{\varsigma}_3 \hat{\varsigma}_1^+ \hat{\varsigma}_2^+ \right) - i\hbar \sum_{j=1}^{3} \kappa_{j,0} \hat{\varsigma}_j^+ \hat{\varsigma}_j. \tag{6}$$

Its first two terms describe the $\chi^{(2)}$ nonlinear interaction with interaction strength $\xi$ [16], while the last non-Hermitian term [37] accounts for intrinsic cavity loss with dissipation rate $\kappa_{j,0} = \omega_j / 2Q_{j,0}$ [22,37], and $Q_{j,0}$ is the intrinsic $Q$ of the microring resonator. Applying the total Hamiltonian in Eq. (3) to two-photon and single-photon input states yields the scattering solutions in coordinate representation. Thereafter, using the commutation relation between momentum $p_x$ and coordinate $x$, $[x, p_x] = i\hbar$ [22,26,38], these solutions can be transformed into momentum space, and hence into the frequency domain, giving the scattering matrices $S_{\text{PUC}}(p_3; k_1, k_2)$ for TPPUC and $S_{\text{PDC}}(p_1, p_2; k_3)$ for SPPDC, where $k_j$ and $p_j$ represent the momentum of input photons and output photons, respectively [36]. Subsequently, the generated photon spectra $\phi_{\text{out, PUC}}(p_3)$ and photon pair joint spectral amplitude (JSA) $\phi_{\text{out, PDC}}(p_1, p_2)$ [10,39] for the TPPUC and SPPDC processes can be obtained according to the input photon state and the scattering matrix, respectively:

$$\phi_{\text{out, PUC}}(p_3) = \iint \mathrm{d}k_1 \mathrm{d}k_2 \phi_{\text{in}}(k_1, k_2) S_{\text{PUC}}(p_3; k_1, k_2), \tag{7}$$

$$\phi_{\text{out, PDC}}(p_1, p_2) = \int \mathrm{d}k_3 \phi_{\text{in}}(k_3) S_{\text{PDC}}(p_1, p_2; k_3), \tag{8}$$

where $\phi_{\text{in}}(k_1, k_2)$ and $\phi_{\text{in}}(k_3)$ are the input JSA of the photon pair for TPPUC and the input spectrum of the single photon for SPPDC, respectively. Finally, the conversion efficiency of the up-converted single photon and the down-converted photon pair can be calculated by integration in the momentum space,

$$\eta_{\text{PUC}} = \int_{-\infty}^{\infty} \mathrm{d}p_3 \left|\phi_{\text{out, PUC}}(p_3)\right|^2, \tag{9}$$

$$\eta_{\text{PDC}} = \int_{-\infty}^{\infty}\int_{-\infty}^{\infty} \mathrm{d}p_1 \mathrm{d}p_2 \left|\phi_{\text{out, PDC}}(p_1, p_2)\right|^2, \tag{10}$$

where $\left|\phi_{\text{out, PUC}}(p_3)\right|^2$ and $\left|\phi_{\text{out, PDC}}(p_1, p_2)\right|^2$ are the probability density functions of the state [10,39]. Hence the complete $\chi^{(2)}$ photon-photon interaction model in the microring-bus system is established.

Because the phase-induced fidelity limitation is removed at the level of the gate principle, the remaining performance metrics of interest are the overall success probability and the purity of the output state. These quantities are determined by the cascading efficiencies of the TPPUC and SPPDC processes and by the spectral correlations introduced in the final down-conversion stage.

We first optimize the TPPUC stage toward maximum conversion efficiency $\eta_{\text{PUC}}$. Specifically, we consider an illustrative microring structure example with a minimized radius of 5 μm, waveguide top width of 600 nm, thickness of 600 nm, and base angle of 80°, with undercut configuration [40-43] to minimize the bending loss. The structure is designed to satisfy the QPM condition 648.55 nm → 1286.63 nm + 1307.74 nm [36]. We consider two representative lineshapes for the uncorrelated input photon pair, namely Gaussian and Lorentzian, as shown in Fig. 4(a). Then we characterize the relation between the $\eta_{\text{PUC}}$ and the linewidth ratio of input photon and the microring $\Gamma$, with fixed $Q$ factors of 650-nm band and 1310-nm band, i.e. $Q_{3,\text{PUC}}$ and $Q_{1,\text{PUC}}$ respectively, as illustrated in Fig. 4(b), showing a clear trade-off. When the input linewidth is much narrower than the cavity linewidth (e.g., $\Gamma$ = 0.1), the temporal duration of the photons becomes long, and the overlap of the two-photon interaction within the cavity lifetime is strongly reduced, leading to a small $\eta_{\rm PUC}$. It appears to contradict the intuition that a narrower photon spectrum corresponds to higher efficiency of entering the microring. Increasing $\Gamma$ improves the temporal overlap, but excessive spectral broadening reduces the coupling efficiency of frequency components that deviate from the cavity resonances. We further find that, for the same bandwidth, Gaussian input spectra yield higher $\eta_{\rm PUC}$ than Lorentzian input spectra because a larger fraction of the spectral weight is concentrated near the cavity resonance. Therefore, we select the Gaussian lineshape for the input photons and subsequently optimize $\eta_{\rm PUC}$ through fine adjustments in $\Gamma$, $Q_{3,\rm PUC}$, and $Q_{1,\rm PUC}$, as shown in Fig. 4(c). Under optimized conditions, we obtain $\eta_{\rm PUC} = 86.0\%$ for $Q_{3,\rm PUC} = 2.15\times10^7$, $Q_{1,\rm PUC} = 2.75\times10^7$ and $\Gamma$ = 1.5. Furthermore, it is worth noting that $\eta_{\rm PUC}$ does not reach unity is itself physically instructive. In case of an uncorrelated two-photon input, different spectral components of a single photon cannot all couple into a microring with identical efficiency. As a result, unit-efficiency TPPUC is unattainable in this configuration even in the presence of strong cavity-enhanced nonlinear interaction.

We next optimize the SPPDC stage to maximize the overall gate success probability $\eta_{\rm CNOT}$ [36], with the results shown in Fig. 4(d). In contrast to TPPUC, the single-photon input to SPPDC does not require temporal overlap with another photon. A narrower input spectrum therefore improves the coupling efficiency into the cavity modes. However, for a fixed input spectrum, this also corresponds to broader cavity linewidth and hence a lower $Q$ factor, which weakens the nonlinear enhancement in the conversion process. This trade-off yields an optimized gate success probability of $\eta_{\rm CNOT}$ = 84.1% for $Q_{3,\rm PDC} = 2.45\times10^7$ and $Q_{1,\rm PDC} = 2.50\times10^7$. Importantly, $\eta_{\rm CNOT}$ remains above 80% over a wide range of $Q_{3,\rm PDC}$ and $Q_{1,\rm PDC}$, indicating that the up-converted photon generated by TPPUC can be converted back with consistently high efficiency. This favorable behavior originates from the close spectral matching between the up-converted single-photon output and the scattering response of the SPPDC stage. The required $Q$ factors for both TPPUC and SPPDC are on the order of $10^7$, which lies well within experimental feasible TFLN capabilities [44,45].

In addition to success probability, the output-state purity is determined by the spectral correlations generated in the SPPDC process. From the calculated up-converted single-photon spectrum $\phi_{\text{out, PUC}}(p_3)$ and the down-converted two-photon

JSA $\phi_{\text{out, PDC}}(p_1, p_2)$ shown in Fig. 4(e) and (f) respectively, we obtain an output-photon purity of 92.9% without pre-compensation. Because the PC-WDM stage allows control of the modulus and phase of the up-converted photon, the two-photon JSA produced in the subsequent SPPDC process can be further engineered toward a less correlated form. As a demonstrative example, we invert the phase distribution of the up-converted single photon, which partially compensates the phase variation introduced by the SPPDC microring and reduces the joint phase correlation in the final two-photon output. The resulting joint phase distribution is shown in Fig. 4(g), and the corresponding output-photon purity increases to 93.1% [36]. Taken together, these results show that the proposed architecture can simultaneously deliver high-success-probability operation, high-purity outputs, and the removal of the intrinsic fidelity ceiling associated with the conventional nonlinear controlled-phase schemes, all within a realistic integrated photonic platform.

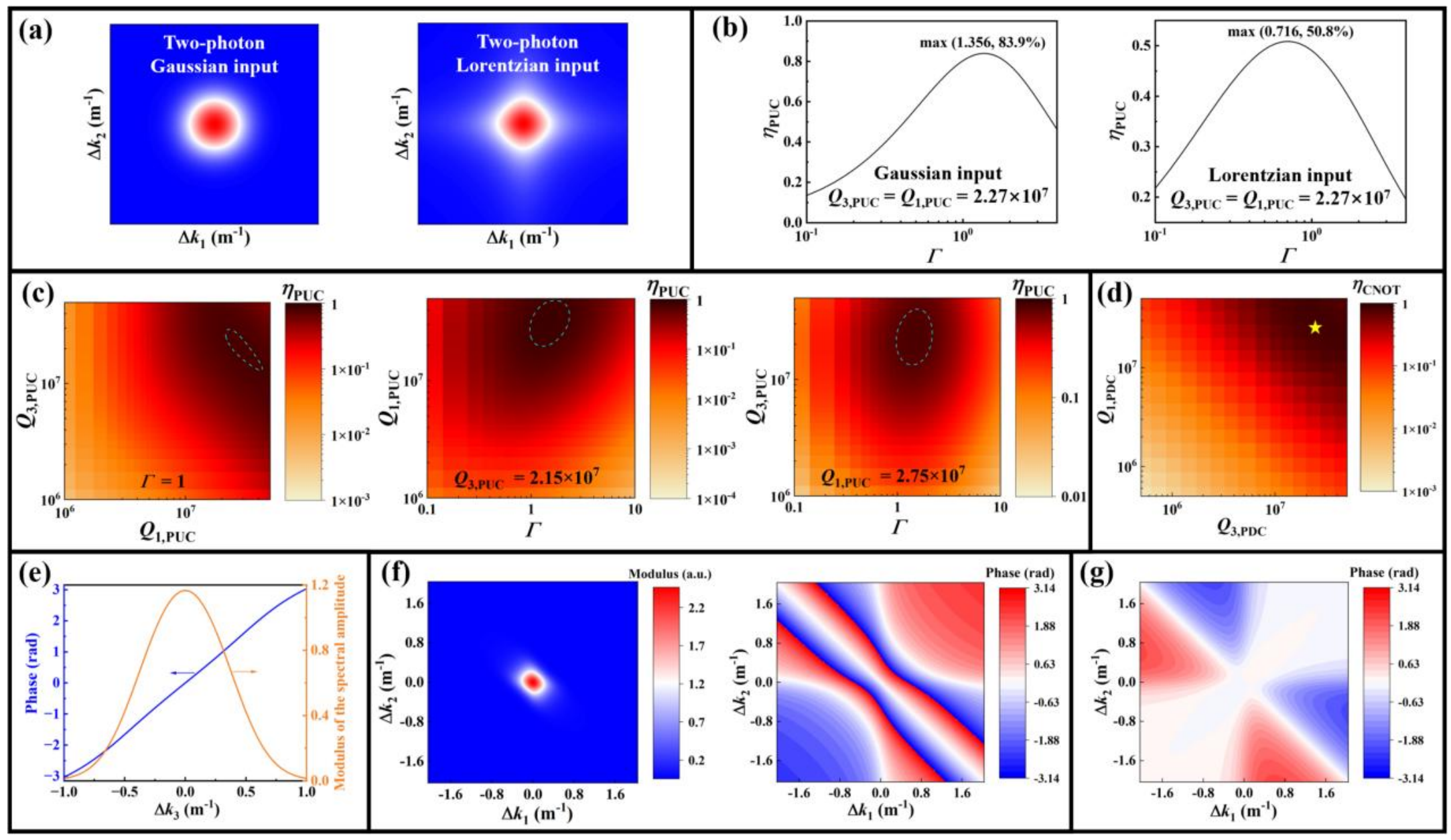

Fig. 4. (a) Uncorrelated two-photon input spectra with Gaussian and Lorentzian lineshapes. (b) The relation between $\eta_{\text{PUC}}$ and $\Gamma$, for two different lineshapes. With $Q_{3,\text{PUC}}$ and $Q_{1,\text{PUC}}$ both fixed at $2.27\times10^7$, a microring with these parameters is capable of achieving deterministic SPPDC [36]. (c) Heatmaps of $\eta_{\text{PUC}}$ versus $Q_{3,\text{PUC}}$, $Q_{1,\text{PUC}}$ and $\Gamma$, fixing two of the parameters and adjusting the other one, with the blue dotted curves marking $\eta_{\text{PUC}} = 80\%$. (d) The heatmap of $\eta_{\text{CNOT}}$ adjusting $Q_{3,\text{PDC}}$ and $Q_{1,\text{PDC}}$, while fixing the optimized PUC parameters of $Q_{3,\text{PUC}} = 2.15\times10^7$, $Q_{1,\text{PUC}} = 2.75\times10^7$ and $\Gamma = 1.5$. (e) The single-photon phase and modulus distribution after TPPUC, under the optimized parameters. (f) The JSA of the photon pair generated by SPPDC without pre-compensation, shown in terms of its modulus and phase. (g) The joint phase distribution of the SPPDC output when the inverted phase profile of the up-converted photon is applied as pre-compensation.

In conclusion, we have proposed an on-chip nonlinear CNOT gate that achieves high success probability, high fidelity, and high-purity outputs targets simultaneously. By decoupling the logical operation from the nonlinear interaction and incorporating spectral-phase pre-compensation, the scheme removes the intrinsic fidelity limitation of conventional controlled-phase gates in principle and improves the output-state quality for the following operations. Guided by a complete quantitative model of cavity-mediated photon-photon interaction, a feasible circuit design is proposed based on the TFLN platform with requirement of $10^7$-$Q$ microresonator, reaching an overall success probability of 84.1%, an output-photon purity of 93.1%, and unity fidelity in principle. This work provides a specific scheme for the scalable CNOT gate, where the remaining work is to optimize the fabrication of the TFLN devices and their integration to authentically for the experimental demonstration. Beyond this specific gate, our work provides a general framework for scalable nonlinear optical logic on chip, and opens a realistic route toward larger-scale photonic quantum circuits based on integrated nonlinear platforms for real quantum computing tasks. Besides quantum computing, such nonlinear photon-photon interaction scheme might also be used for many other applications, such as generation of single-photon state [34,46], $N$-photon states [16], quantum metrology [47,48], etc., which is a powerful building block for the overall quantum information technology.

**Acknowledgments**

This work was supported by National Key R&D Program of China (No.2025YFF0524600), National Natural Science Foundation of China (62305156, 62293523, 62293520), the fundamental Research funds for the central Universities (021014380250), Postgraduate Education Reform Project of Jiangsu Province (2025JGZD090), and Fundamental and Interdisciplinary Disciplines Breakthrough Plan of the Ministry of Education of China (JYB2025XDXM106). We thank Prof. Y. Y. Ren, Dr. M. X. Li, Dr. Z. Y. Tao and Dr. B. Wu for helpful discussions.

**Code and Data Availability**

The code and data that support the findings of this study are available from the corresponding author upon reasonable request.

## Author Contributions

Z. X., H-Y. L. and M. S. conceived the original idea, M. S. and H-Y. L. conceived and designed the research. M. S. developed the theoretical framework, M. S., Y. W., X. L. and Q. N. conducted the calculation and simulation. M. S. and H-Y. L. performed the figure organization and prepared the manuscript, H-Y. L. and Z. X. supervised the project. All authors revised the final manuscript.

## Competing Interests

The authors declare no competing interests.

## References:

[1] Harrow, A.W. and A. Montanaro, Nature **549**, 203-209 (2017).

[2] Zhong, H.-S., et al., Science **370**, 1460-1463 (2020).

[3] Chabaud, U. and M. Walschaers, Physical Review Letters **130**, 090602 (2023).

[4] Gottesman, D., Physical Review A **57**, 127 (1998).

[5] Fowler, A.G., M. Mariantoni, J.M. Martinis, and A.N. Cleland, Physical Review A **86**, 032324 (2012).

[6] Wang, D.S., A.G. Fowler, and L.C. Hollenberg, Physical Review A **83**, 020302 (2011).

[7] Noiri, A., et al., Nature **601**, 338-342 (2022).

[8] Xue, X., et al., Nature **601**, 343-347 (2022).

[9] Kok, P., et al., Reviews of Modern Physics **79**, 135-174 (2007).

[10] Liu, Y.-C., et al., Scientific Reports **11**, 12628 (2021).

[11] Knill, E., R. Laflamme, and G.J. Milburn, Nature **409**, 46-52 (2001).

[12] O'Brien, J.L., et al., Nature **426**, 264-267 (2003).

[13] Langford, N.K., et al., Nature **478**, 360-363 (2011).

[14] Niu, M.Y., I.L. Chuang, and J.H. Shapiro, Physical Review Letters **120**, 160502 (2018).

[15] Yanagimoto, R., et al., Optica **9**, 1289-1296 (2022).

[16] Liu, H.-Y., et al., Advanced Photonics Nexus **2**, 016003 (2023).

[17] Xie, Z. and S. Zhu, Advanced Photonics **4**, 030502 (2022).

[18] Xie, Z., et al., Advances in Physics: X **9**, 2322739 (2024).

[19] Mobini, E., D.H. Espinosa, K. Vyas, and K. Dolgaleva, Micromachines **13**, 991 (2022).

[20] Zhao, M. and K. Fang, Optica **9**, 258-263 (2022).

[21] Akin, J., et al., Light: Science & Applications **13**, 290 (2024).

[22] Li, M., et al., Physical Review Applied **13**, 044013 (2020).

[23] Burridge, B.M., I.I. Faruque, J.G. Rarity, and J. Barreto, Optica **10**, 1471-1477 (2023).

[24] Zhu, D., et al., Advances in Optics and Photonics **13**, 242-352 (2021).

[25] Chen, G., et al., Advanced Photonics **4**, 034003 (2022).

[26] Lee, C., C. Noh, N. Schetakis, and D.G. Angelakis, Physical Review A **92**, 063817 (2015).

[27] Zheng, H., D.J. Gauthier, and H.U. Baranger, Physical Review A **82**, 063816 (2010).

[28] Shen, J.-T. and S. Fan, Physical Review Letters **98**, 153003 (2007).

[29] Shen, J.-T. and S. Fan, Physical Review A **76**, 062709 (2007).
[30] Rephaeli, E. and S. Fan, Physical Review Letters **108**, 143602 (2012).
[31] Yuan, R., et al., Chinese Physics Letters **42,** 070401 (2025).
[32] Yu, Y., et al., ACS Photonics **9**, 3253-3259 (2022).
[33] Ma, Z., et al., Physical Review Letters **125**, 263602 (2020).
[34] Chen, J.-Y., et al., Physical Review Applied **16**, 064004 (2021).
[35] Zhu, D., et al., Light: Science & Applications **11**, 327 (2022).
[36] See Supplemental Material for details.
[37] Scully, M.O. and M.S. Zubairy, *Quantum Optics*. 1997: Cambridge University Press.
[38] Shen, J.-t. and S. Fan, Optics Letters **30**, 2001-2003 (2005).
[39] Ansari, V., et al., Physical Review Letters **120**, 213601 (2018).
[40] Xie, R.R., G.Q. Li, F. Chen, and G.L. Long, Advanced Optical Materials **9**, 2100539 (2021).
[41] Lu, R. and S. Gong, Journal of Micromechanics and Microengineering **31**, 114001 (2021).
[42] Liu, X., et al., Optics Letters **45**, 6318-6321 (2020).
[43] Liang, H., et al., Optica **4**, 1251-1258 (2017).
[44] Gao, R., et al., New Journal of Physics **23**, 123027 (2021).
[45] Zhang, M., et al., Nature Photonics **13**, 36-40 (2019).
[46] Wang, H., et al., Laser & Photonics Reviews **19**, 2401420 (2025).
[47] Krischek, R., et al., Physical Review Letters **107**, 080504 (2011).
[48] Barbieri, M., PRX Quantum **3**, 010202 (2022).